\documentclass[twocolumn, amssymb, aps, prl]{revtex4}

\usepackage{graphicx}
\usepackage{dcolumn}
\usepackage{bm}
\usepackage{amsmath}
\usepackage{latexsym}
\usepackage{float}
\usepackage{amssymb}
\usepackage{graphicx}
\usepackage{epsfig}
\usepackage{psfrag}
\usepackage{amsmath}
\usepackage{textcomp}
\usepackage{hyperref}
\begin{document}

{\bf \noindent Comment on ``Mode-coupling Theory as a Mean-Field Description of the Glass Transition''}

\bigskip

Motivated by the interpretation of mode-coupling theory (MCT) as a
mean-field theory Ikeda and Miyazaki (IM) \cite{1} and
the present authors \cite{2} independently
investigated the dependence of MCT for hard spheres, with diameter
$\sigma$ on the spatial dimension $d$, particularly in the limit
$d \rightarrow \infty$. A comparison with the corresponding results
from replica theory \cite{3} has revealed serious discrepancies
between both theories \cite{1,2}. The long time
limit of the self part of the van Hove function $G_{c, \infty}^{(s)} (r;d)$ is the Fourier
transform of the critical self-nonergodicity parameter (NEP)
$f^{(s)}_c(k;d)$:
\begin{equation} \label{eq1}
G_{c, \infty}^{(s)} (r;d) =A_d\,  r^{-(d/2-1)} \int\limits_0^\infty
dk\, k^{d/2} J_{d/2-1} (kr) f^{(s)}_c (k;d) \quad
\end{equation}
($A_d=(2\pi)^{-d/2}$). Taking $f^{(s)}_c (k;d)$ from MCT, IM show that
$r^{d-1} G_{c, \infty}^{(s)} (r;d)$
has negative dips on a scale $r/\sigma=O(1)$ for $d=4, \cdots, 15$,
contradicting its non-negativity.
From these results IM conclude that a
``reconsideration and revision of MCT from ground up is in
order.'' In the following we will explain why these observations
are not yet sufficient to draw that conclusion.

\begin{figure}[h]
\centerline{%
\includegraphics[width=6.5cm,clip=true]{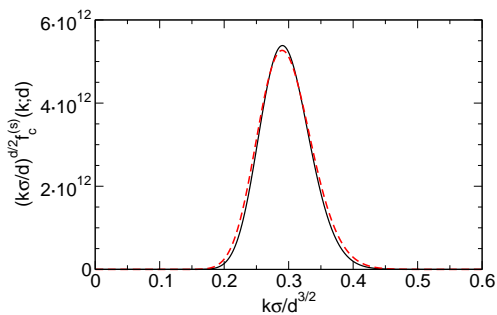}}
\caption{(colour online) $k$ dependence of $(k\sigma /d)^{d/2}f^{(s)}_c(k;d)$ (solid line)
from the numerical MCT solution and $(\sigma /d)^{d/2}\bar{g}^{(s)}_c(k;d)$
(dashed line) for $d=100$.}
\label{fig1}
\end{figure}

First we show that the dips in $G_{c, \infty}^{(s)} (r;d)$ may disappear for $d\rightarrow \infty$.
From our numerical approach we have found that
$k^{d/2}f^{(s)}_c(k;d)\rightarrow \bar{g}^{(s)}_c(k;d)=k^{d/2}\exp [-a_{d}(k-k_0)/\sqrt{d}]$
($a_{\infty}\cong1.50\sigma$, $k_0\cong0.155\sigma^{-1}d^{3/2}$) for $d\rightarrow\infty$
(see Fig.~\ref{fig1} for $d=100$).

The numerical result for $G_{c, \infty}^{(s)}(r;100)$ (not shown) exhibits
tiny negative dips. However, replacing $k^{d/2}f^{(s)}_c(k;d)$ in Eq.~(\ref{eq1})
by $\bar{g}^{(s)}_c(k;d)$ for $d\gg1$ one obtains the analytical result
$G_{c, \infty}^{(s)}(r;d)\rightarrow \bar{G}_{c, \infty}^{(s)}(r;d)=B_d/[1+(r/a_{d})^2 d]^{(d+1)/2}$ 
($0<B_d\sim d^d$). Hence $G_{c, \infty}^{(s)}(r;d)$ becomes non-negative
(and a Gaussian on a scale $rd/\sigma=O(1)$) for $d\rightarrow \infty$.
The same holds for the corresponding collective quantity $G_{c, \infty}(r;d)$.
For $d=100$ this also demonstrates that a small deformation of $k^{d/2}f^{(s)}_c(k;d)$ can
already eliminate the negative dips even for a non-Gaussian $f^{(s)}_c(k;d)$.

A second comment is inspired by the investigation of a
mean-field $\phi^4$-model on a lattice with $N$ sites \cite{5}.
It has been shown that a time scale $\tau(N) \sim N$ exists.
The dynamics is nonergodic for
$t\ll\tau(N)$ and ergodic for $t \gg \tau(N)$ for \textit{all} temperatures
\cite{5}. In case that such a diverging time scale $\tau(d)$ would exist for a
$d$-dimensional liquid, as well, both limits $t \rightarrow
\infty$ and $d \rightarrow \infty$ would {\it not} commute. This would suggest
to investigate, e.g., the self-correlator $S^{(s)}$ in a
more general scaling limit
$\hat{S}^{(s)} (\hat{k}, \hat{t}; \hat{\varphi})=\lim\limits_{d
\rightarrow \infty} S^{(s)} (d^\rho \hat{k}/ \sigma, d^\eta \hat{t};
d^\kappa 2^{-d} \hat{\varphi}; d)$
with appropriate scaling exponents $\rho, \eta$ and $\kappa$.
This has not been done so far. In Refs. \cite{1,2} where $\eta=0$
is assumed the limit
$t \rightarrow \infty$ has been taken first for large but {\it
finite} $d$.

The non-negativity of $G_{c, \infty}^{(s)}(r;d)$ and $G_{c, \infty}(r;d)$
for $d\rightarrow\infty$ does not imply that MCT becomes exact for
$d\rightarrow \infty$. If it would turn out that MCT  does not become exact for
$d\rightarrow\infty$ this would imply that MCT is not a mean-field
theory, at least in the conventional
sense. This holds indeed for the $\phi^4$ model with $\eta =0$ \cite{5}.
Concerning $d=O(1)$, we fully agree with the existence of negative dips in
$r^{d-1}G_{c, \infty}^{(s)}(r;d)$. Of course, one has to take into
account that they are already strongly
enhanced  for $d=15$ due to the factor $r^{d-1}$ included by IM.
Using PY theory we found that they also exist in $d=2$ and $3$ for
$G_{c, \infty}^{(s)}(r;d)$ but not for $G_{c, \infty}(r;d)$.
Because of the \textit{nonlinear} structure of MCT equations this is
not surprising and does not seem to affect the quality of MCT
successfully tested particularly for three-dimensional liquids during more than
two decades. Of course, a deeper insight, based on the removal of the negative
dips or not, why MCT is so powerful in $d=3$ and even $d=2$ and whether it is a
kind of mean-field theory or not is highly desirable.\\

\noindent Rolf Schilling and Bernhard Schmid \\
\noindent Institut f\"ur Physik, Johannes
Gutenberg-Universit\"at\\
\noindent D-55099 Mainz, Germany\\



\vspace{-0.5cm}

\begin{thebibliography}{99}

\bibitem{1} A. Ikeda and K. Miyazaki, Phys. Rev. Lett. {\bf 104},
255704 (2010)

\bibitem{2} B. Schmid and R. Schilling, Phys. Rev. E {\bf81},
041502 (2010)

\bibitem{3} G. Parisi and F. Zamponi, J. Stat. Mech.: Theory Exp
(2006) P03017.

\bibitem{5} W. Kob and R. Schilling, J. Phys.: Condens. Matter
{\bf3}, 9195 (1991)

\end{thebibliography}
\end{document}